\documentclass[runningheads]{llncs}
\usepackage[T1]{fontenc}
\usepackage{graphicx}
\usepackage{comment}% Added for comments
\usepackage{pgfplots}% Added for figures
\pgfplotsset{compat=1.16}
\usepackage{xurl}
\usepackage{xcolor}
\usepackage{booktabs}
\usepackage{cite}
\usepackage{hyperref}
\begin{document}
\title{Towards Real-time Intrahepatic Vessel Identification in
Intraoperative Ultrasound-Guided Liver Surgery}
\titlerunning{Towards Real-time Intrahepatic Vessel Identification in
IOUS}

%%% \author{Author\inst{1}\orcidID{0000-1111-2222-3333}

%\author{Author\inst{1,2} \and Author\inst{2} \and Author\inst{1} \and Author\inst{1}}

\author{
Karl-Philippe Beaudet\inst{1, 2, 3} \and 
Alexandros Karargyris\inst{1,2} \and 
Sidaty El Hadramy\inst{2,3} \and 
Stéphane Cotin\inst{2,3} \and 
Jean-Paul Mazellier\inst{1,2} \and 
Nicolas Padoy\inst{1,2} \and 
Juan Verde\inst{1,2,3}}
%\author{
%Karl-Philippe Beaudet\inst{1, 3}\orcidID{0009-0003-4417-546X} \and 
%Alexandros Karargyris\inst{1,2}\orcidID{0000-0002-1930-3410} \and 
%Sidaty El Hadramy\inst{2,3}\orcidID{0009-0000-2917-0706} \and 
%Stéphane Cotin\inst{2,3}\orcidID{0000-0002-2661-505X} \and 
%Jean-Paul Mazellier\inst{1,2}\orcidID{0000-0003-1659-0642} \and 
%Nicolas Padoy\inst{1,2}\orcidID{0000-0002-5010-4137} \and 
%Juan Verde\inst{1,2,3}\orcidID{0000-0002-9127-8467}}
%index{Beaudet, Karl-Philippe}
%index{Karargyris, Alexandros}
%index{El Hadramy, Sidaty}
%index{Cotin, Stéphane}
%index{Mazellier, Jean-Paul}
%index{Padoy, Nicolas}
%index{Verde, Juan}

%
\authorrunning{K.-P. Beaudet et al.}
\institute{IHU Strasbourg, Strasbourg, France \and University of Strasbourg, CNRS, INSERM, ICube, UMR7357, Strasbourg, France \and Inria, Strasbourg, France}
\maketitle              % typeset the header of the contribution
\begin{abstract}

While laparoscopic liver resection is less prone to complications and maintains patient outcomes compared to traditional open surgery, its complexity hinders widespread adoption due to challenges in representing the liver's internal structure. Laparoscopic intraoperative ultrasound offers efficient, cost-effective and radiation-free guidance. Our objective is to aid physicians in identifying internal liver structures using laparoscopic intraoperative ultrasound. We propose a patient-specific approach using preoperative 3D ultrasound liver volume to train a deep learning model for real-time identification of portal tree and branch structures. Our personalized AI model, validated on \textit{ex vivo} swine livers, achieved superior precision (0.95) and recall (0.93) compared to surgeons, laying groundwork for precise vessel identification in ultrasound-based liver resection. Its adaptability and potential clinical impact promise to advance surgical interventions and improve patient care.

\end{abstract}
\section{Introduction}\label{sec1}

Primary liver cancer ranks seventh globally among all cancer types, representing 8.2\% of cancer-related deaths in 2018 \cite{bray2018global}. Liver resection remains the primary potentially curative method, with laparoscopic surgery showing promise in reducing postoperative complications \cite{fretland2018laparoscopic}. However, challenges persist, including the risk of unintended damage to intrahepatic vascular structures \cite{ciria2016comparative}. 
Intraoperative ultrasound (IOUS) has emerged as a valuable tool for guiding surgery, particularly in assessing tumor margins and avoiding vascular damage \cite{el2023trackerless, hagopian2014abdominal}. However, IOUS faces challenges such as restricted workspace and the need for multiple exchanges between instruments and the ultrasound probe \cite{hagopian2020liver}.
Laparoscopic IOUS (LIOUS) introduces further complexity as it necessitates a reinterpretation of US images in the longitudinal-like plane, demanding a process of relearning and retraining, as well as a limited range of movement within the abdominal cavity \cite{hagopian2014abdominal}. These limitations underscore the need for innovative solutions. 
We propose leveraging machine learning to enhance the efficacy and safety of laparoscopic liver resection. However, the scarcity and quality of training datasets pose significant challenges, compounded by the operator-dependent nature of ultrasound (US) imaging. 
Offline annotation of 2D US video streams is laborious and unreliable, particularly with laparoscopic IOUS.

Image guidance systems (IGS) are applied to laparoscopic liver resection (LLR) to enhance safety and intraoperative orientation \cite{kingham2013evolution, nicolau2011augmented}. IGS allows surgeons to visualize structures like tumors and blood vessels from preoperative scans that are not visible with a laparoscope \cite{nicolau2011augmented}. However, laparoscopic ultrasound is limited by its 2D nature and poor tumor contrast \cite{kingham2013evolution}. 
Vessel segmentation in US is explored for 3D B-Mode volumes with adaptive thresholding and Hessian post-processing \cite{nam2012automatic}. However, there are no commercially available 3D laparoscopic US probes.
For 2D US in liver imaging, a segmentation method using Hessian-based filters and shape elimination is proposed in  \cite{song2015locally}.
Deep learning with convolutional neural networks is also increasingly used for 2D US hepatic vasculature segmentation. U-Net methods offer high segmentation performance with 2D and 3D US datasets \cite{ronneberger2015u}. Other authors propose to use a U-Net for LUS vessel segmentation to support automatic untracked LUS to CT registration \cite{ramalhinho2020registration, montana2021vessel}. However US-to-CT registration faces challenges like slow inference times, demanding faster and more efficient registration algorithms for real-time integration of preoperative imaging data. 
Finally, some authors advocate using generative adversarial networks (GANs) \cite{zaman2020generative} or physics-inspired methods \cite{tirindelli2021rethinking} to construct datasets for training ultrasound segmentation models. However, existing studies primarily address simpler applications like bone surface segmentation, while our research focuses on identifying hepatic vessels.

This paper introduces contributions aimed at surmounting the challenges in LLR and improving preoperative-to-intraoperative translation. In the conventional LLR workflow, preoperative CT scans are performed for surgical planning and intraoperative guidance. However, surgeons are often required to mentally translate preoperative 3D data into 2D LIOUS. In order to provide a solution for accurate intrahepatic vascular structure identification in LIOUS, our contributions entail proposing a clinically applicable modified workflow consisting in (1) acquiring a 3D US liver volume from preoperative 2D US liver scan sequences; (2) semi-automatically segmenting the intrahepatic vascular structures in the 3D volume; (3) training a patient-specific deep learning model from the pre-operative volume to identify portal branches in intraoperative US in real-time. To train the patient specific model, we propose to use a dataset of synthetic 2D US images along with their corresponding segmentation obtained by reslicing the 3D US volume. The reslicing is combined with data augmentation to enhance the generalization capabilities of the intraoperative artificial intelligence (AI) model.

By providing surgeons with real-time guidance and precise vascular identification, our system could potentially improve liver cancer surgery, benefiting both patients and healthcare providers.
The paper is organized as follows: Sect. \ref{Methods} details the method and its novelty, Sect. \ref{Sec3} presents our current results on \textit{ex vivo} porcine data, and ﬁnally, we conclude in Sect. \ref{Sec4} and discuss future work.

\section{Methods} \label{Methods}

\begin{figure}
\centering
\includegraphics[width=1.00\textwidth]{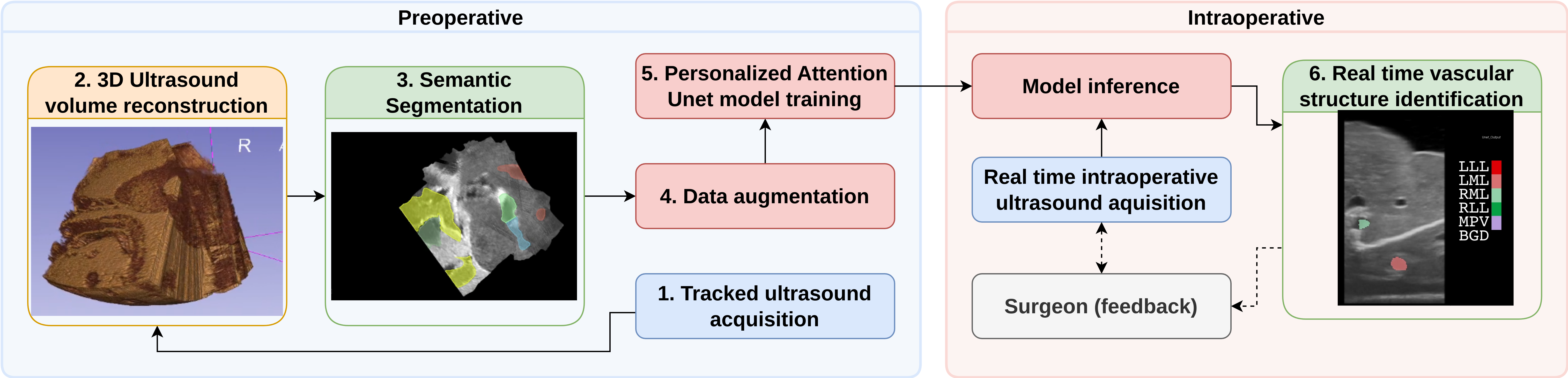}
\caption{Overview of the proposed method for real-time portal branch identification in the liver. The method comprises 6 steps: 1) Electromagnetic (EM) tracked US acquisition, 2) 3D US volume reconstruction, 3) Semantic segmentation of the reconstructed volume, 4) Data augmentation to create a personalized dataset for diverse US scanning protocols, 5) Personalized model training using an Attention U-Net (A-UNet) architecture \cite{oktay2018attention}, and 6) Real-time deployment of the model using 3D Slicer \cite{fedorov20123d}.}\label{figMethodOverview}
\end{figure}

In our work, we simplified several problems to deliver a proof of concept while retaining core constraints. Firstly, we assumed the structural integrity of the \textit{ex vivo} swine liver during imaging, using a rigid gel-based scaffold to prevent deformation. Secondly, we maintained a fixed ultrasound depth of 9cm to encompass the liver model adequately. Lastly, we focused on developing a personalized, patient-specific identification model rather than a generalized one, aligning with practicality and surgical workflow integration. This involved preoperative liver scans to gather patient-specific data for model training, potentially using a 3D ultrasound liver volume for intraoperative inference via a laparoscopic probe. These considerations form the basis of our real-time liver portal branch identification method, detailed in six steps in Fig. \ref{figMethodOverview}. Subsequent sections will delve into each step.

\textbf{1. Tracked US acquisition}: 
We utilized a US system with a linear transducer to scan the liver (see supplementary materials Sect. 1), mimicking a preoperative US scan. The pose of the transducer was tracked using an EM tracking system, allowing us to record 2D tracked US sequences. Notably, the linear probe, chosen deliberately as a worst-case scenario, required multiple scans, potentially resulting in gaps or overlaps between swipes.

\textbf{2. 3D US volume reconstruction}: 
We employed the volume reconstruction algorithm from \cite{lasso2014plus} to generate 3D US volumes from 2D tracked US sequences. This algorithm redistributes tracked US frames into the 3D volume space through linear interpolation, with voxel values determined as weighted averages of overlapping pixels. To prevent gaps, especially in high-resolution volumes, we applied a hole-filling algorithm that interpolates voxel values using a spherical Gaussian kernel based on nearby known voxels.

\begin{figure}
\centering
\includegraphics[width=1.00\textwidth]{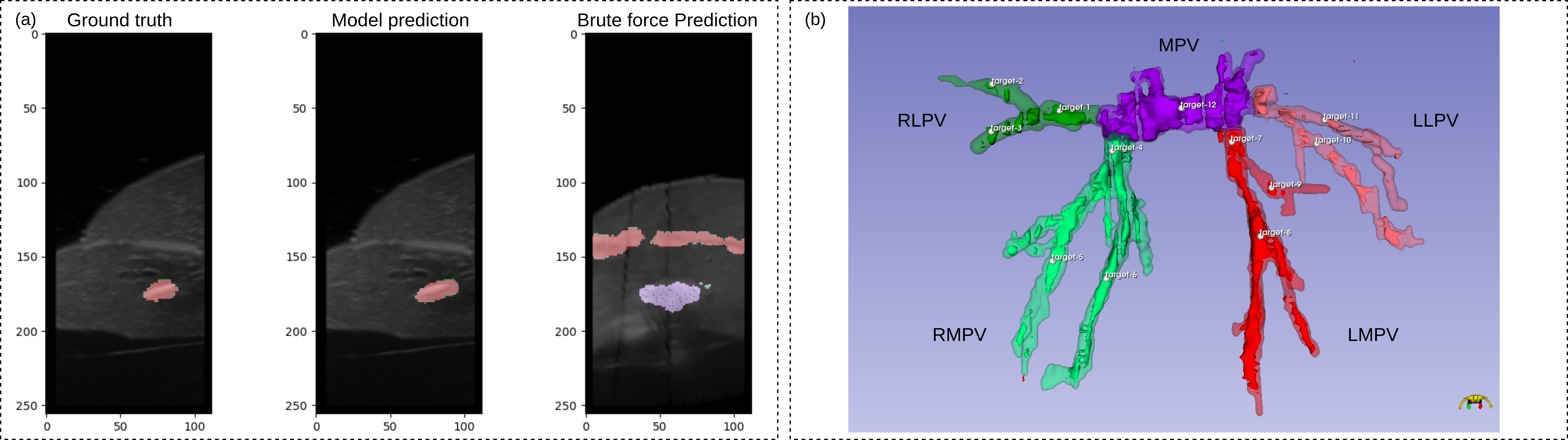}
\caption{Validation experiment comparing ground truth, model prediction, and brute force method (model 2). \textbf{(a)} Ground truth projection on a US frame, model prediction, and brute force method overlay. \textbf{(b)} Overlay of AI model predictions (opaque) and 3D segmentation ground truth (transparent) from reconstructed 3D US volume.}
\label{figValidation}
\end{figure}

\textbf{3. Semantic segmentation}: The reconstructed US volume was annotated by a senior surgeon (see Fig. \ref{figValidation}.b), who was asked to segment only "large" vessels confidently identifiable, omitting "small" vessels (diameter $\le$ 2mm) due to clinical irrelevance and visibility/connectivity limitations. The surgeon segmented the key hepatic vessels (adapted to swine anatomy), including the Main Portal Vein (MPV), the Right Lateral Portal Vein (RLPV), the Right Medial Portal Vein (RMPV), the Left Medial Portal Vein (LMPV), and the Left Lateral Portal Vein (LLPV) (see Fig. \ref{figValidation}.b).

\textbf{4. Data augmentation (Reslicing)}: 
We employed a semi-supervised data augmentation pipeline to generate labeled synthetic 2D US images from segmented 3D US volumes. The reslicing, based on spline interpolation, incorporated intensity variation for improved generalization across different scanning protocols. Realistic US maneuvers such as tilting, rocking, sliding, transversal sliding, and lifting were applied to adapt to user and scanning protocols. Validation with a clinician ensured the synthetic images closely resembled realistic US scans. Central cropping was performed according to linear dimensions and aspect ratio. This pipeline allowed us to create a personalized dataset of segmented synthetic 2D US images by reslicing each 3D US volume.

\textbf{5. Personalized model training}: 
We trained a patient-specific 2D attention U-Net model \cite{oktay2018attention} (model choice justification in Sect. \ref{modeljustification}) on the labeled synthetic 2D US image (containing all the possible US scanning protocols) to learn the hepatic labels in the synthetic 2D US images.

\textbf{6. AI-enabled intraoperative assistance}:
Our personalized 2D segmentation model enabled intraoperative support by identifying intrahepatic vessels. Real-time inference on unseen 2D US sequences or real-time US acquisition was performed using a linear transducer, corresponding to the typical US guidance in LLR. In laparoscopic US, similar signal dimensions to linear transducers were observed despite potential intensity variations from tissue contact. Our data augmentation compensated for contrast and gain differences, ensuring minimal impact on Unet performance and maintaining consistent size and shape of image features across transducers.

\section{Results} \label{Sec3}

\subsection{Dataset and Implementation Details}

\begin{itemize}
    \item \textbf{Tracked US sequence and scanning technique}:
    We utilized an ACUSON S3000 HELX Touch US system with a linear probe (9L4) set to a 9 cm depth for a full transversal view of the liver. An EM tracking system (trakSTAR, NDI, Canada) with a 6 DoF sensor attached to the US transducer was employed. US images were captured using a frame grabber (AV.io HD+, Epiphan Video), and tracked sequences were recorded using the PLUS toolkit and 3D Slicer. Calibration for spatial and temporal alignment was achieved using the Freehand tracked US calibration application (fCal) from the PLUS toolkit. The frame size was 450 × 188 pixels with an image spacing of 0.5 mm per pixel.
    Validation was performed on 2 \textit{ex vivo} swine livers. Tracked US sequences were obtained by scanning the entire liver surface at 15 fps in approximately 2 minutes, resulting in 1446 and 3016 images per sequence for each \textit{ex vivo} liver model, respectively.
    \item \textbf{Volume reconstruction}:
    The volume were reconstructed from the tracked US sequence using the 3D Slicer IGT algorithm from PLUS toolkit \cite{lasso2014plus, fedorov20123d}. An output spacing of 0.5 mm/pixel was chosen to preserve a good resolution. The reconstructed 3D US volumes had a size of 550x450x150 pixels.
    \item \textbf{Annotations}:
    The 3D US volumes (2 livers) were annotated with 5 labels (multi labels annotations) which corresponded to the MPV (32.29\%, 32.72\%), the RLPV (9.92\%, 9.63\%), RMPV (23.72\%, 28.1\%), the LMPV (26.63\%, 17.5\%) and the LLPV (7.44\%, 12.05\%).
    \item \textbf{Data augmentation parameters}:
    We adapted to user and US scanning protocols by reslicing 3D US volumes with rotations (±180°), tilting (±30°), rocking (±12°), sliding/sweeping (80\%), transversal sliding (40\%), and pressure/lifting (15\%). Central cropping was based on linear US dimensions and aspect ratio (9 cm depth, 0.418 aspect ratio), with intensity augmentations replicating variations in US gain (±30\% brightness and ±30\% contrast), along with random horizontal flips. This pipeline generated a personalized dataset of 50,000 annotated synthetic 2D US images for each \textit{ex vivo} liver model (see Fig. \ref{figMethodOverview}). Source code and data will be made publicly available at \url{https://github.com/CAMMA-public/Lupin}.
    \item \textbf{Model architecture and training}:
    We utilized a 2D Attention U-Net architecture with a depth of 1024, trained from scratch, with an image input size of 256x112 and 6 output labels for segmentation. The data for each \textit{ex vivo} liver was split into training and validation sets with a ratio of 9:1, while test sets consisted of test US sequences. Our method was implemented in PyTorch\footnote{https://pytorch.org/docs/stable/index.html.} 1.7.1 and the MONAI\footnote{https://docs.monai.io/en/stable/index.html.} 1.2.0 framework, trained and evaluated on a GeForce RTX 6000 GPU. We employed an Adam optimizer with a learning rate of $5 \times 10^{-5}$ and a warm restart learning rate scheduler. The loss function combined DICE and cross-entropy (1:1 ratio) with Adam optimizer. Training converged in 20 epochs with a batch size of 10. The model with the best performance on the validation data was selected for testing. Hyperparameters considered included batch size, learning rate (and scheduler), image size, number of segmentation channels, DICE and cross-entropy loss ratio, and model size (number of filters).
    \item \textbf{System testers cohort}:
    The cohort included 4 surgeons with varying levels of expertise (junior to senior) in liver procedures and US imaging methods.
    
\end{itemize}
\subsection{Model validation} \label{modeljustification}

\textbf{Metrics and protocol}
We evaluated our model's intrahepatic structure identification using the DICE score, a standard metric for segmentation tasks \cite{moccia2018blood}, emphasizing the importance of segmentation alongside identification accuracy for visual representation.
We compared the model's 2D segmentation predictions to the 3D segmentation ground truth projected onto initial 2D US frames. Model validation involved running inference on tracked US sequences and computing the DICE score to assess segmentation accuracy. Additionally, we used a brute force method for comparison, which identified the closest US image from a dataset of 50,000 images using the Structural Similarity Metric \cite{wang2004image}. This served as a baseline to justify using a neural network. Real-time capabilities were evaluated by measuring frames per second (fps) processed in the test sequences.

\begin{figure}
\centering
\includegraphics[width=0.75\textwidth]{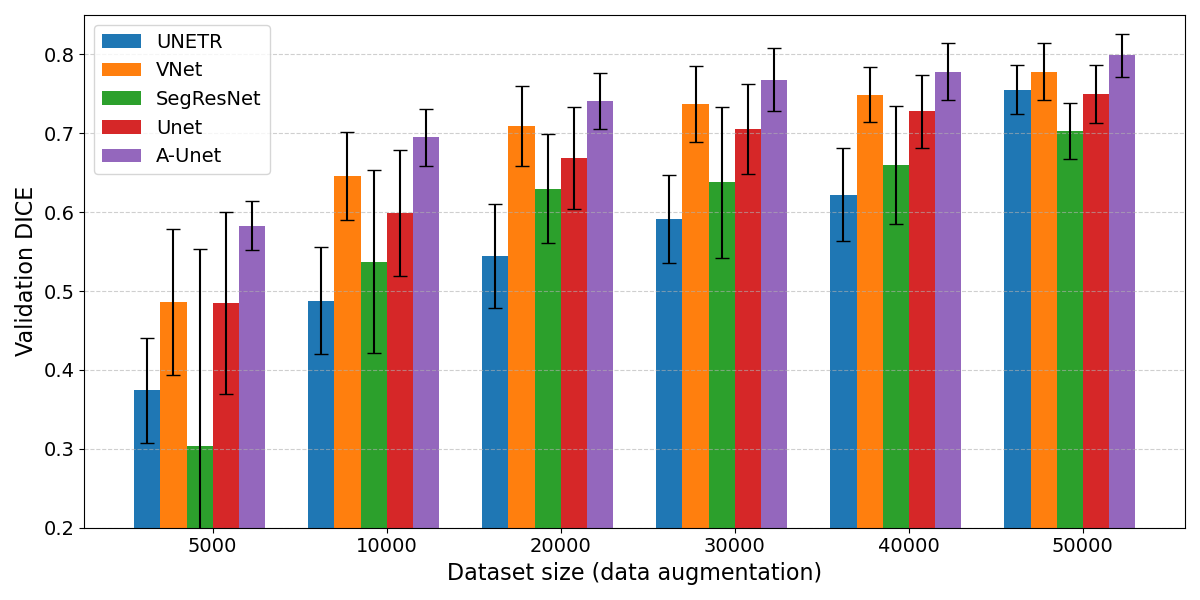}
\caption{Comparison model architectures for the identification of intrahepatic vascular structures and the effect of data augmentation}\label{graphAblationSensitivity}
\end{figure}

\textbf{Ablation and sensibility study}
We evaluated several model architectures: U-Net \cite{falk2019u}, Attention U-Net \cite{oktay2018attention}, SegResNet \cite{myronenko20193d}, VNet \cite{milletari2016v}, and UNETR \cite{hatamizadeh2021unetr}. Using the Validation DICE score as the primary metric, Attention U-Net outperformed all other architectures across various dataset sizes, achieving a score of 0.799 for a dataset size of 50,000 (see Fig. \ref{graphAblationSensitivity}). Therefore, we selected Attention U-Net as the optimal architecture for our hepatic vascular structure identification task due to its highest Validation DICE score.
To evaluate the impact of clinical data augmentation on model generalization across different US protocols and knobology settings, we examined how training dataset size affects model performance measured by the DICE metric. Modifying the dataset size allowed us to adjust the diversity of parameter variations present within the training data. Results of this sensitivity analysis are shown in Fig. \ref{graphAblationSensitivity}.

\textbf{Results} \label{subsubModelResults}
Our method outperformed the brute force approach in vessel identification and localization accuracy on validation sequences. With an inference time of $0.072 \pm 0.002$ s ($13.89 \pm 0.39$ fps) compared to the brute force method's $18.42 \pm 0.01$ s ($0.05429 \pm 0.0001$ fps) for a 50,000 images dataset, our method enables real-time identification of vascular structures in IOUS. The mean DICE score for our method on the US frames test set was $0.60 \pm 0.05$, significantly higher than the brute force approach's $0.030 \pm 0.088$ (see table \ref{tabModelVal}).

\begin{table}[ht]
\centering
\caption{DICE score of each sequences of each model per branch type with an A-UNet architecture and a dataset of 50,000 images}\label{tabModelVal}
\begin{tabular}{ccccccc}
\toprule
Liver & MPV & RLPV & RMPV & LMPV & LLPV & Mean \\
\midrule
1 & $0.67\pm0.23$ & $0.35\pm0.30$ & $0.60\pm0.20$ & $0.60\pm0.27$ & $0.71\pm0.29$ & $0.59\pm0.04$ \\
2 & $0.66\pm0.20$ & $0.44\pm0.30$ & $0.71\pm0.19$ & $0.68\pm0.17$ & $0.53\pm0.22$ & $0.61\pm0.05$ \\
%3 & $0.57\pm0.33$ & $0.43\pm0.30$ & $0.59\pm0.26$ & $0.69\pm0.24$ & $0.67\pm0.23$ & $0.59\pm0.04$ \\
\bottomrule
\end{tabular}
\end{table}

\subsection{Pre-clinical validation}

\textbf{Metrics and protocol}
To validate our solution in a clinical setting, four surgeons identified 12 targets in the portal branches. Surgeons selected optimal views, communicated identifications, and markers were placed accordingly. Surgeons had unlimited time and were aided by 3D segmentation. The AI model was trained exclusively on portal labels for a fair comparison.
Performance assessment compared the model to surgeons by evaluating spatial portal branch identification. Surgeons annotated US frames in real-time, while the model performed inference on those frames. Precision was employed as the metric for both the model and surgeons. Recall was computed for the model, not for surgeons, due to predetermined target identifications. True positives were determined based on the model's predictions and surgeons' annotations within the ground truth volume.
The validation tracked US sequence was registered to the 3D volume, allowing projection of segmentation onto 2D frames. A \textit{5 mm} error tolerance was allowed for annotation predictions \cite{ruiter2006model}. Performance was compared for two tasks: distinguishing portal and hepatic veins and identifying portal vein branches.

\textbf{Results}
We compared surgeons' performance with the Attention U-Net model in portal branch identification, as summarized in Table \ref{tabSurgeonsVsModel}. Possible sources of bias include differences in orientation between the ground truth and surgeons' perspectives and limited anatomical context in our \textit{ex vivo} liver models. However, these biases are consistent across our model and experimental conditions. The model generally outperforms surgeons, achieving a mean precision of 0.95 and recall of 0.93, compared to surgeons' mean precision ranging from $0.50$ to $0.86$ for portal vein tree identification and from $0.05$ to $0.54$ for portal vein branch identification (see Table \ref{tabSurgeonsVsModel}). Model identification errors primarily consist of false negatives due to recall, but this is not a significant issue for our application, as precision is prioritized to minimize false positives. Surgeons' lower identification scores can be attributed to misidentification of the portal vein leading to complete misidentification of its branches. Surgeons's feedback on our real-time hepatic vascular structures identification tool (see supplementary materials Sect. 2) was generally positive, with mean scores ranging from 3.75 to 4.75 on a scale of 0 (Strongly Disagree) to 5 (Strongly Agree), indicating high satisfaction, confidence in identification, and educational value of the tool (see Table \ref{tabSurgeonsFeedback}).

\begin{table}[ht]
\caption{Comparison of Surgeons and Attention U-Net in Portal Branch identification tasks: Portal Vein Tree (PVT) and Portal Vein Branches (PVB) identification}\label{tabSurgeonsVsModel}
\begin{tabular*}{\textwidth}{@{\extracolsep\fill}lcccccccc}
\toprule
 & & Surgeon 1 & Surgeon 2 & Surgeon 3 & Surgeon 4 & \multicolumn{2}{c}{A-Unet} \\
\cmidrule{7-8}
ID & LVR & \multicolumn{4}{c}{Precision (PPV)} & PPV & TPR \\
\toprule    
PVT & 1 & $0.63\pm0.52$ & $0.22\pm0.40$ & $1.00\pm0.00$ & $0.71\pm0.49$ & $1.00\pm0.00$ & $0.94\pm0.20$ \\
       & 2 & $0.67\pm0.49$ & $1.00\pm0.00$ & $0.00\pm0.00$ & $1.00\pm0.00$ & $1.00\pm0.00$ & $0.91\pm0.19$ \\
       & Mean & $0.65\pm0.51$ & $0.61\pm0.10$ & $0.50\pm0.00$ & $0.86\pm0.25$ & $1.00\pm0.00$ & $0.93\pm0.20$ \\
\midrule
PVB & 1 & $0.13\pm0.35$ & $0.00\pm0.00$ & $0.10\pm0.32$ & $0.57\pm0.53$ & $0.94\pm0.13$ & $0.94\pm0.20$ \\
       & 2 & $0.50\pm0.45$ & $0.67\pm0.50$ & $0.00\pm0.00$ & $0.50\pm0.53$ & $0.96\pm0.11$ & $0.91\pm0.19$ \\
       & Mean & $0.32\pm0.40$ & $0.34\pm0.25$ & $0.05\pm0.16$ & $0.54\pm0.53$ & $0.95\pm0.12$ & $0.93\pm0.20$ \\
\bottomrule
\end{tabular*}
\end{table}

\begin{table}[ht]
\centering
\caption{Overview of surgeons (SG) feedback form using the hepatic vascular structures identification tool (where 1 represents "Strongly Disagree" and 5 "Strongly Agree")}\label{tabSurgeonsFeedback}%
\begin{tabular}{@{}llllll@{}}
\toprule
Question & SG 1  & SG 2 & SG 3 & SG 4 & Mean \\
\midrule
1. Intraoperative tool effectiveness & 4 & 2 & 5 & 5 & $4.00\pm1.22$ \\
%2. Tool ease of use & 5 & 2 & 5 & 5 & $4.25\pm1.30$ \\
%3. Visualization clarity & 5 & 2 & 5 & 4 & $4.00\pm1.22$ \\
2. Confidence in identification & 4 & 2 & 4 & 5 & $3.75\pm1.09$ \\
3. Tool's value & 5 & 2 & 5 & 5 & $4.25\pm1.30$ \\
%6. No Complications/Adverse events & 5 & 5 & 5 & 5 & $5.00\pm0.00$ \\
4. Workflow integration & 4 & 2 & 5 & 4 & $3.75\pm1.09$ \\
5. Real-time feedback & 5 & 2 & 5 & 5 & $4.25\pm1.30$ \\
%9. Recommend to other surgeons & 5 & 1 & 5 & 5 & $4.00\pm1.73$ \\
6. Educational potential & 5 & 4 & 5 & 5 & $4.75\pm1.43$ \\
\bottomrule
\end{tabular}
\end{table}

\section{Conclusion}\label{Sec4}

In this study, we introduced a novel method for vessel identification in intraoperative ultrasound frames. Our approach combines tracked preoperative ultrasound acquisition, volume reconstruction, data augmentation, and personalized model training. Validation experiments on humanized swine livers showed that our method accurately identifies vessels, surpassing surgeon performance with high precision and low false discovery rates. Its adaptability to various scanning protocols and knobology enhances clinical applicability, while personalized model training improves accuracy. Additionally, streamlining the training process with advanced 3D segmentation tools could potentially lead to full automation. Future work will focus on enhancing the data augmentation pipeline to handle deformations and incorporating temporal modeling for further improvements.

\begin{credits}
\subsubsection{\ackname} This work has been partially funded under the framework of the French Investments for the Future Program, by French state funds managed within the "Plan Investissements d'Avenir", by the ANR (reference ANR-10-IAHU-02) and by the Interdisciplinary Thematic Institute HealthTech, as part of the ITI 2021-2028 program of the University of Strasbourg, CNRS, and Inserm, supported by IdEx Unistra (ANR-10-IDEX-0002) and SFRI (STRAT'US project, ANR-20-SFRI-0012). The authors would like to express their gratitude to the Preclinical Research Unit of the IHU Strasbourg for their invaluable support and assistance throughout this study.

\subsubsection{\discintname} The authors have no competing interests to declare that are relevant to the content of this article.
\end{credits}

\newpage
\bibliographystyle{splncs04}
\bibliography{paper_4218}

\begin{thebibliography}{10}
\providecommand{\url}[1]{\texttt{#1}}
\providecommand{\urlprefix}{URL }
\providecommand{\doi}[1]{https://doi.org/#1}

\bibitem{bray2018global}
Bray, F., Ferlay, J., Soerjomataram, I., Siegel, R.L., Torre, L.A., Jemal, A.: Global cancer statistics 2018: {{GLOBOCAN}} estimates of incidence and mortality worldwide for 36 cancers in 185 countries. CA  \textbf{68}(6),  394--424 (2018). \doi{10.3322/caac.21492}

\bibitem{ciria2016comparative}
Ciria, R., Cherqui, D., Geller, D.A., Briceno, J., Wakabayashi, G.: Comparative {{Short-term Benefits}} of {{Laparoscopic Liver Resection}}: 9000 {{Cases}} and {{Climbing}}. Annals of Surgery  \textbf{263}(4),  761--777 (Apr 2016). \doi{10.1097/SLA.0000000000001413}

\bibitem{el2023trackerless}
El~Hadramy, S., Verde, J., Beaudet, K.P., Padoy, N., Cotin, S.: Trackerless volume reconstruction from intraoperative ultrasound images. In: MICCAI. pp. 303--312. Springer (2023)

\bibitem{falk2019u}
Falk, T., Mai, D., Bensch, R., {\c{C}}i{\c{c}}ek, {\"O}., Abdulkadir, A., Marrakchi, Y., B{\"o}hm, A., Deubner, J., J{\"a}ckel, Z., Seiwald, K., et~al.: U-net: deep learning for cell counting, detection, and morphometry. Nature methods  \textbf{16}(1),  67--70 (2019)

\bibitem{fedorov20123d}
Fedorov, A., Beichel, R., Kalpathy-Cramer, J., Finet, J., Fillion-Robin, J.C., Pujol, S., Bauer, C., Jennings, D., Fennessy, F., Sonka, M., et~al.: 3d slicer as an image computing platform for the quantitative imaging network. Magnetic resonance imaging  \textbf{30}(9),  1323--1341 (2012), \url {https://www.slicer.org/} (visited: 2023-06-02)

\bibitem{fretland2018laparoscopic}
Fretland, {\AA}.A., Dagenborg, V.J., Bj{\o}rnelv, G.M.W., Kazaryan, A.M., Kristiansen, R., Fagerland, M.W., Hausken, J., T{\o}nnessen, T.I., Abildgaard, A., Barkhatov, L., Yaqub, S., R{\o}sok, B.I., Bj{\o}rnbeth, B.A., Andersen, M.H., Flatmark, K., Aas, E., Edwin, B.: Laparoscopic {{Versus Open Resection}} for {{Colorectal Liver Metastases}}: {{The OSLO-COMET Randomized Controlled Trial}}. Annals of Surgery  \textbf{267}(2),  199--207 (Feb 2018). \doi{10.1097/SLA.0000000000002353}

\bibitem{hagopian2020liver}
Hagopian, E.J.: Liver ultrasound: A key procedure in the surgeon's toolbox. Journal of surgical oncology  \textbf{122}(1),  61--69 (2020)

\bibitem{hagopian2014abdominal}
Hagopian, E.J., Machi, J. (eds.): Abdominal {{Ultrasound}} for {{Surgeons}}. {Springer New York}, {New York, NY} (2014). \doi{10.1007/978-1-4614-9599-4}

\bibitem{hatamizadeh2021unetr}
Hatamizadeh, A., Tang, Y., Nath, V., Yang, D., Myronenko, A., Landman, B., Roth, H., Xu, D.: Unetr: Transformers for 3d medical image segmentation. arxiv. arXiv preprint arXiv:2103.10504  (2021)

\bibitem{kingham2013evolution}
Kingham, T.P., Jayaraman, S., Clements, L.W., Scherer, M.A., Stefansic, J.D., Jarnagin, W.R.: Evolution of image-guided liver surgery: transition from open to laparoscopic procedures. Journal of Gastrointestinal Surgery  \textbf{17},  1274--1282 (2013)

\bibitem{lasso2014plus}
Lasso, A., Heffter, T., Rankin, A., Pinter, C., Ungi, T., Fichtinger, G.: Plus: open-source toolkit for us-guided intervention systems. IEEE Trans. Biomed. Eng.  \textbf{61}(10),  2527--2537 (2014)

\bibitem{milletari2016v}
Milletari, F., Navab, N., Ahmadi, S.A.: V-net: Fully convolutional neural networks for volumetric medical image segmentation. In: 2016 fourth international conference on 3D vision (3DV). pp. 565--571. Ieee (2016)

\bibitem{moccia2018blood}
Moccia, S., De~Momi, E., El~Hadji, S., Mattos, L.S.: Blood vessel segmentation algorithms—review of methods, datasets and evaluation metrics. Computer methods and programs in biomedicine  \textbf{158},  71--91 (2018)

\bibitem{montana2021vessel}
Monta{\~n}a-Brown, N., Ramalhinho, J., Allam, M., Davidson, B., Hu, Y., Clarkson, M.J.: Vessel segmentation for automatic registration of untracked laparoscopic ultrasound to ct of the liver. IJCARS  \textbf{16}(7),  1151--1160 (2021)

\bibitem{myronenko20193d}
Myronenko, A.: 3d mri brain tumor segmentation using autoencoder regularization. In: BrainLesion: Glioma, Multiple Sclerosis, Stroke, and Traumatic Brain Injuries (4th International Workshop). pp. 311--320. Springer (2019)

\bibitem{nam2012automatic}
Nam, W.H., Kang, D.G., Lee, D., Lee, J.Y., Ra, J.B.: Automatic registration between 3d intra-operative ultrasound and pre-operative ct images of the liver based on robust edge matching. Physics in Medicine \& Biology  \textbf{57}(1),  69--91 (2012)

\bibitem{nicolau2011augmented}
Nicolau, S., Soler, L., Mutter, D., Marescaux, J.: Augmented reality in laparoscopic surgical oncology. Surgical oncology  \textbf{20}(3),  189--201 (2011)

\bibitem{oktay2018attention}
Oktay, O., Schlemper, J., Folgoc, L.L., Lee, M., Heinrich, M., Misawa, K., Mori, K., McDonagh, S., Hammerla, N.Y., Kainz, B., et~al.: Attention u-net: Learning where to look for the pancreas. arXiv preprint arXiv:1804.03999  (2018)

\bibitem{ramalhinho2020registration}
Ramalhinho, J., Tregidgo, H.F., Gurusamy, K., Hawkes, D.J., Davidson, B., Clarkson, M.J.: Registration of untracked 2d laparoscopic ultrasound to ct images of the liver using multi-labelled content-based image retrieval. IEEE Transactions on Medical Imaging  \textbf{40}(3),  1042--1054 (2020)

\bibitem{ronneberger2015u}
Ronneberger, O., Fischer, P., Brox, T.: U-net: Convolutional networks for biomedical image segmentation. In: MICCAI. pp. 234--241. Springer (2015)

\bibitem{ruiter2006model}
Ruiter, N.V., Stotzka, R., Muller, T.O., Gemmeke, H., Reichenbach, J.R., Kaiser, W.A.: Model-based registration of x-ray mammograms and mr images of the female breast. IEEE Transactions on Nuclear Science  \textbf{53}(1),  204--211 (2006)

\bibitem{song2015locally}
Song, Y., Totz, J., Thompson, S., Johnsen, S., Barratt, D., Schneider, C., Gurusamy, K., Davidson, B., Ourselin, S., Hawkes, D., et~al.: Locally rigid, vessel-based registration for laparoscopic liver surgery. IJCARS  \textbf{10},  1951--1961 (2015)

\bibitem{tirindelli2021rethinking}
Tirindelli, M., Eilers, C., Simson, W., Paschali, M., Azampour, M.F., Navab, N.: Rethinking ultrasound augmentation: A physics-inspired approach. In: MICCAI. pp. 690--700. Springer (2021)

\bibitem{wang2004image}
Wang, Z., Bovik, A.C., Sheikh, H.R., Simoncelli, E.P.: Image quality assessment: from error visibility to structural similarity. IEEE Trans. Image. Process.  \textbf{13}(4),  600--612 (2004)

\bibitem{zaman2020generative}
Zaman, A., Park, S.H., Bang, H., Park, C.w., Park, I., Joung, S.: Generative approach for data augmentation for deep learning-based bone surface segmentation from ultrasound images. IJCARS  \textbf{15},  931--941 (2020)

\end{thebibliography}
\end{document}

\maketitle

\section{Supplementary Material}

\subsection{\textit{Ex vivo} swine liver model}

\begin{figure}[h]
\centering
\includegraphics[width=1.00\textwidth]{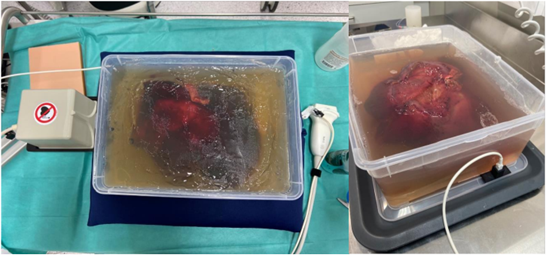}
\caption{Humanized \textit{ex vivo} swine liver model within a stiff, US-compatible, gel-based scaffold}\label{LiverModel}
\end{figure}

This section presents images of the \textit{ex vivo} swine liver model. Figure \ref{LiverModel} illustrates a humanized \textit{ex vivo} swine liver within a stiff, US-compatible, gel-based scaffold. The tracked ultrasound probe, electromagnetic reference sensor of the liver box, and the field emitter of the electromagnetic system are also visible.

\newpage

\subsection{Surgeon's feedback form}

\begin{figure}[h]
\centering
\includegraphics[width=1.00\textwidth]{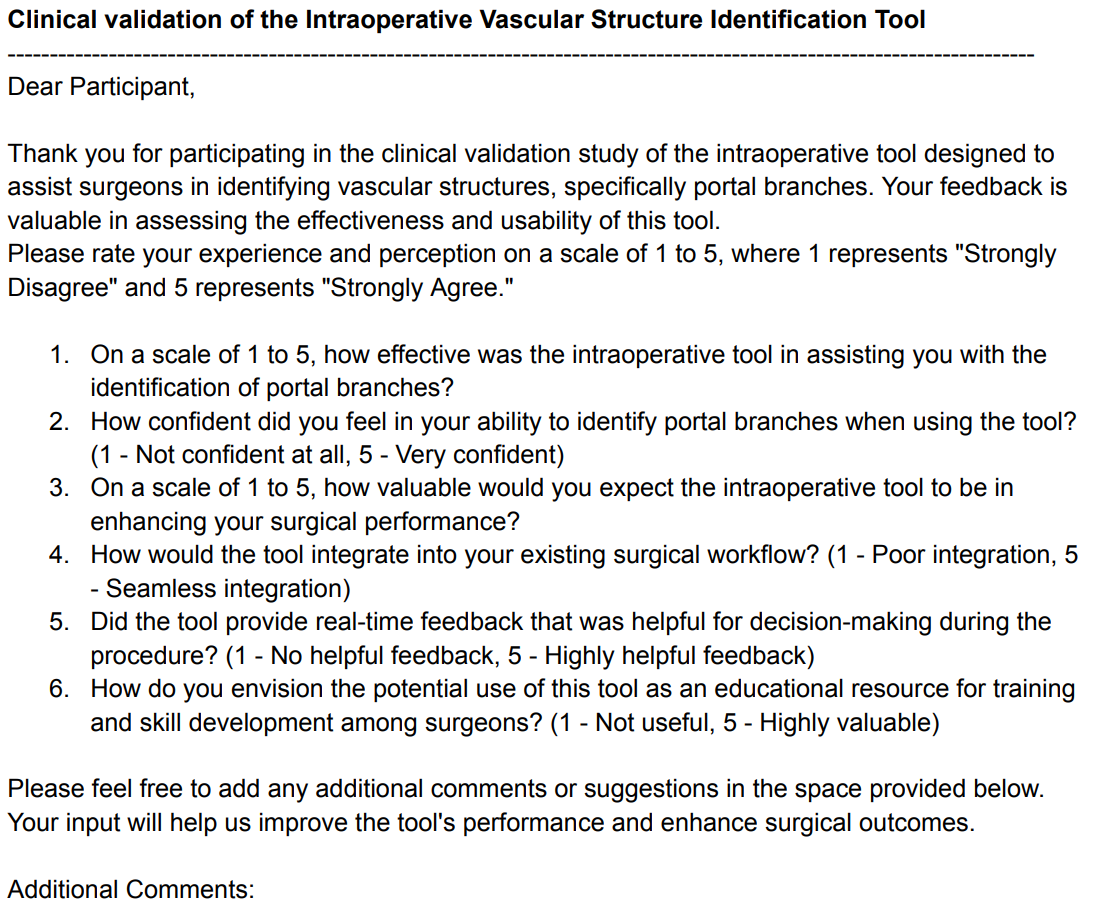}
\caption{Feedback form provided to surgeons after the experiments}\label{FeedbackForm}
\end{figure}

This section provides the feedback form (see Figure \ref{FeedbackForm}) used to collect surgeons' feedback on the Intrahepatic Vessel Identification tool after validation experiments.

